\documentclass[nofootinbib,aps,pra,twocolumn,english]{revtex4-1}
\pdfoutput=1

\usepackage{orcidlink}
\usepackage{amsfonts}

\usepackage{color}
\usepackage{graphicx}
\usepackage{wrapfig,enumerate,slashed}
\usepackage[utf8]{inputenc}
\usepackage{adjustbox}

\usepackage[english]{babel}
\usepackage[T1]{fontenc}
\usepackage{amsmath}
\usepackage{hyperref}
\usepackage{subfig}
\usepackage{caption}

\usepackage{appendix}
\usepackage{placeins}

\usepackage{footmisc}
\usepackage{amsmath}
\usepackage{wasysym} 
\usepackage{graphicx}
\usepackage{color}
\usepackage{comment}
\usepackage{hyperref}
\usepackage{slashed}
\usepackage{booktabs}

\newcommand{\be}{\begin{eqnarray}}
\newcommand{\ee}{\end{eqnarray}}

\newcommand{\bee}{\begin{eqnarray}}
\newcommand{\eee}{\end{eqnarray}}
\newcommand{\beeq}{\begin{equation}}
\newcommand{\eeeq}{\end{equation}}
\newcommand{\fref}[1]{Fig.~\ref{#1}}

\usepackage{listings}
\usepackage{color}
\usepackage{hyperref}

\definecolor{dkgreen}{rgb}{0,0.6,0}
\definecolor{gray}{rgb}{0.5,0.5,0.5}
\definecolor{mauve}{rgb}{0.58,0,0.82}
\definecolor{cyan}{rgb}{0.88,1,1}
\definecolor{TopRow}{rgb}{0.4,0.7,1}
\definecolor{lblue}{rgb}{0.8,0.9,1}

\begin{document}
{\flushright 
\hspace{6.5cm} IPPP/22/66}

\title{A Genetic Quantum Annealing Algorithm}


\author{Steven~A.~Abel \orcidlink{0000-0003-1213-907X} }
\email{steve.abel@durham.ac.uk }
\author{Luca~A.~Nutricati \orcidlink{0000-0002-5045-5113}}
\email{luca.a.nutricati@durham.ac.uk}
\author{Michael Spannowsky
\orcidlink{0000-0002-8362-0576}}
\email{michael.spannowsky@durham.ac.uk}
\affiliation{\vspace{0.1cm} Institute for Particle Physics Phenomenology, Durham University, Durham DH1 3LE, UK}
\affiliation{Department of Mathematical Sciences, Durham University, Durham DH1 3LE, UK}

\begin{abstract}
{\small
A genetic algorithm (GA) is a search-based optimization technique based on the principles of Genetics and Natural Selection. We present an algorithm which enhances the classical GA with input from quantum annealers. As in a classical GA, the algorithm works by breeding a population of possible solutions based on their fitness. However, the population of individuals is defined by the continuous couplings on the quantum annealer, which then give rise via quantum annealing to the set of corresponding phenotypes that represent attempted solutions. This introduces a form of directed mutation into the algorithm that can enhance its performance in various ways. Two crucial enhancements come from the continuous couplings having strengths that are inherited from the fitness of the parents (so-called {\it nepotism}) and from the annealer couplings allowing the entire population to be influenced by the fittest individuals (so-called {\it quantum-polyandry}). We find our algorithm to be significantly more powerful on several simple problems than a classical GA. 
}
\end{abstract}

\maketitle
\flushbottom


\section{\label{Sec:Intro}Introduction}

The purpose of this paper is to present an evolutionary algorithm that enhances the classical genetic algorithm (GA) with input from adiabatic quantum computers, or quantum annealers. 

The general idea of combining quantum computing and evolutionary algorithms is a relatively old one in the framework of traditional gate quantum computers, but the success that is claimed in this arena is on a fairly restricted set of problems (see  \cite{Sofge:2008,computers5040024,Ibarrondo:2022psk,dur36576} for reviews).
The situation with quantum annealers is even less well developed \cite{Chancellor:2016ebx,dur36576,Coxson2014AdiabaticQC,King2019QuantumAssistedGA,Dollen2022QuantumenhancedSO}. However both GAs and quantum annealers 
have each individually been having  increasing impact in similar domains, especially recently in the context of particle physics and string theory (see for example ~\cite{Yamaguchi:1999hq,Allanach:2004my,Akrami:2009hp,Blaback:2013ht,Abel:2014xta,Ruehle:2017mzq,Abel:2018ekz,Cole:2019enn,AbdusSalam:2020ywo,Ruehle:2020jrk,Bena:2020xrh,Larfors:2020ugo,Bena:2021wyr,Abel:2021rrj,Abel:2021ddu,Cole:2021nnt,Loges:2021hvn} and \cite{Abel:2020ebj, Abel:2020qzm, Abel:2021fpn} respectively). This still strongly suggests that there is benefit to be gained by combining them. 

The relatively limited progress in this direction most likely has to do with the conflicting philosophies behind quantum computing and GAs. At first glance one might suppose that GAs must be quite compatible with quantum computing. Indeed they work by evolving a population of individuals that are each typically defined by a binary code (the so-called {\it genotype}) that is easily translatable into qubit eigenvalues. The stumbling block however is the fact that a GA operates by selecting the individuals in the population for breeding based on their fitness (i.e. how close they come to solving the problem of interest). Adopting a purist quantum computing perspective therefore, one would have to encode the entire problem to be solved in the quantum computer. In other words one would have to encode how to get from the binary genotype of an individual to its phenotype and thence to the fitness, using nothing but quantum circuitry. Moreover quantum annealers are not universal, so even if one could encode interesting problems in this fashion, the problems that could be treated by a {\it purely} quantum annealing system  would be restricted to only those that can be expressed as a quadratic Ising model. 

Contrast this with the approach one typically takes when utilising a GA classically. Usually one begins with an already well-defined and possibly very complicated system that needs to be optimised in some way. A GA then performs the desired optimisation  by treating the system itself as a ``black box''. All that is required is a means of calling the  ``black box'' with a binary input (i.e. the genotype), and have it provide an output that can be assessed using a fitness function. This agnostic aspect of GAs is in part what gives them their great power because the system to be optimised can simply be ``bolted on''. The problems that can be treated this way are virtually limitless and unconstrained. From this perspective it is not  clear {\it a priori} if the gain to be made by incorporating GAs into quantum computers {\it in their entirety} outweighs the inevitable loss of this fundamental advantage. For this reason the approach that we wish to advocate in this paper is a hybrid one, in which this crucial advantage of GAs is preserved. We call this hybrid approach a {\it Genetic Quantum Annealing Algorithm} (GQAA).  \vspace{-1cm}

\section{Genetic Quantum Annealing}

In order to describe our method for combining the two techniques, we will first briefly recapitulate the elements behind classical GAs  ~\cite{turing,Holland1975,David1989,Holland1992,Forrest1993,Jones1995,Collard1998,Reeves2002,haupt,Michalewicz2004}. We follow this with a brief overview of quantum annealers, in order to get all the required ingredients in place, and establish our notation, before turning to the mechanism itself. 

\subsection{Genetic algorithms}

GAs work by evolving a population of individuals each defined by a binary code, the aforementioned {\it genotype}. The genotype of each individual yields its particular properties, the so-called {\it phenotype}, which is essentially the list of parameters that define the problem of interest. 

\begin{figure}[!htbp]
 \vspace{-0.75cm}
 \hspace{-0.5cm}
 \begin{center}
\includegraphics[width=.5\textwidth]{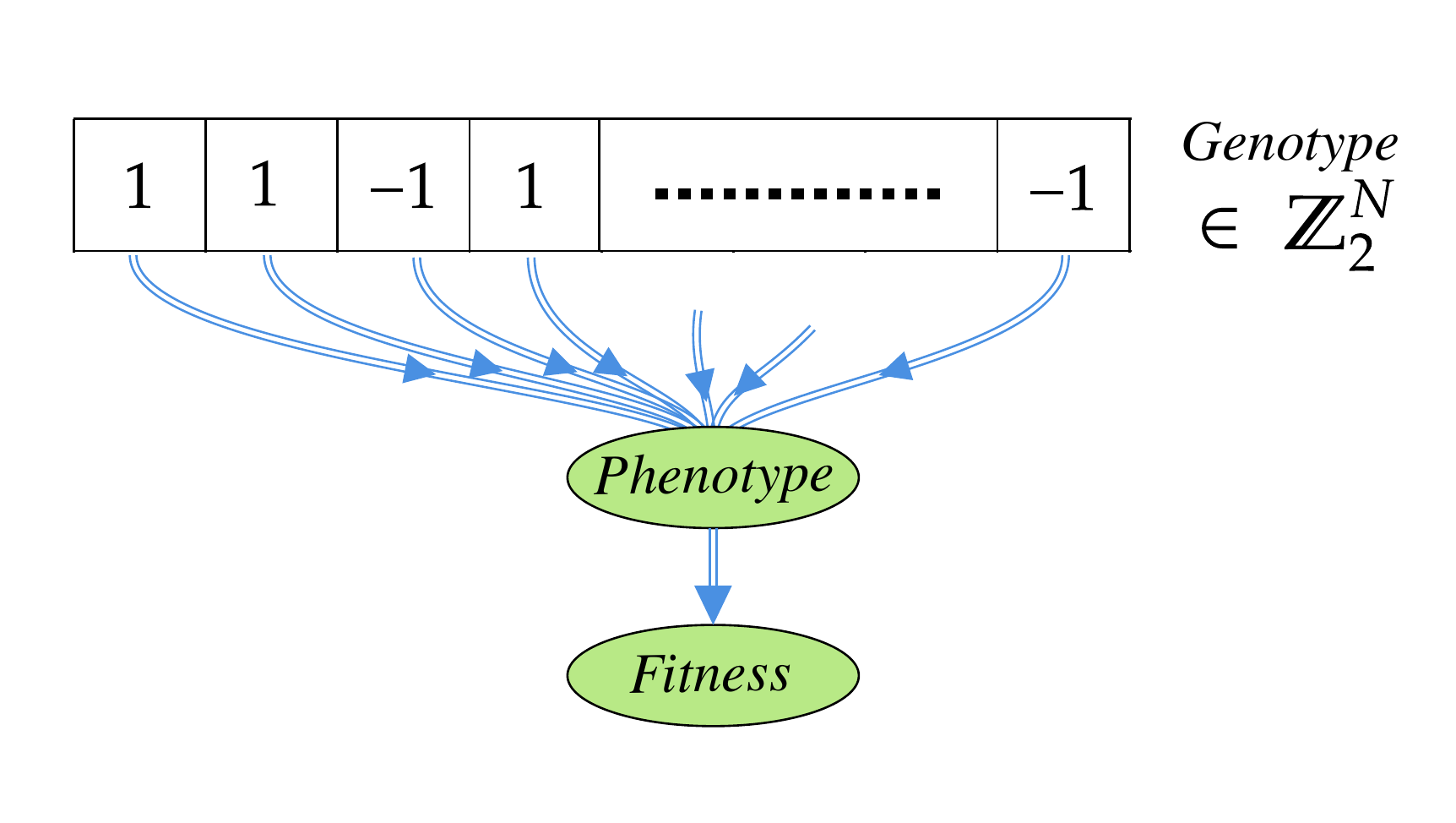}
\caption{An individual member of the population in a classical GA, where here the genotype contains $N$ discrete alleles (for pedagogical purposes we represent the genotype with $\sigma = \pm 1$ spin values, which are translated to binary digits $\tau$ in the obvious way, $\tau=(1+\sigma)/2$). }
\label{fig:individualGA}
\end{center}
\end{figure}

\begin{figure*}[!htbp]
 \vspace{-0.75cm}
 \begin{center}
\includegraphics[width=.85\textwidth]{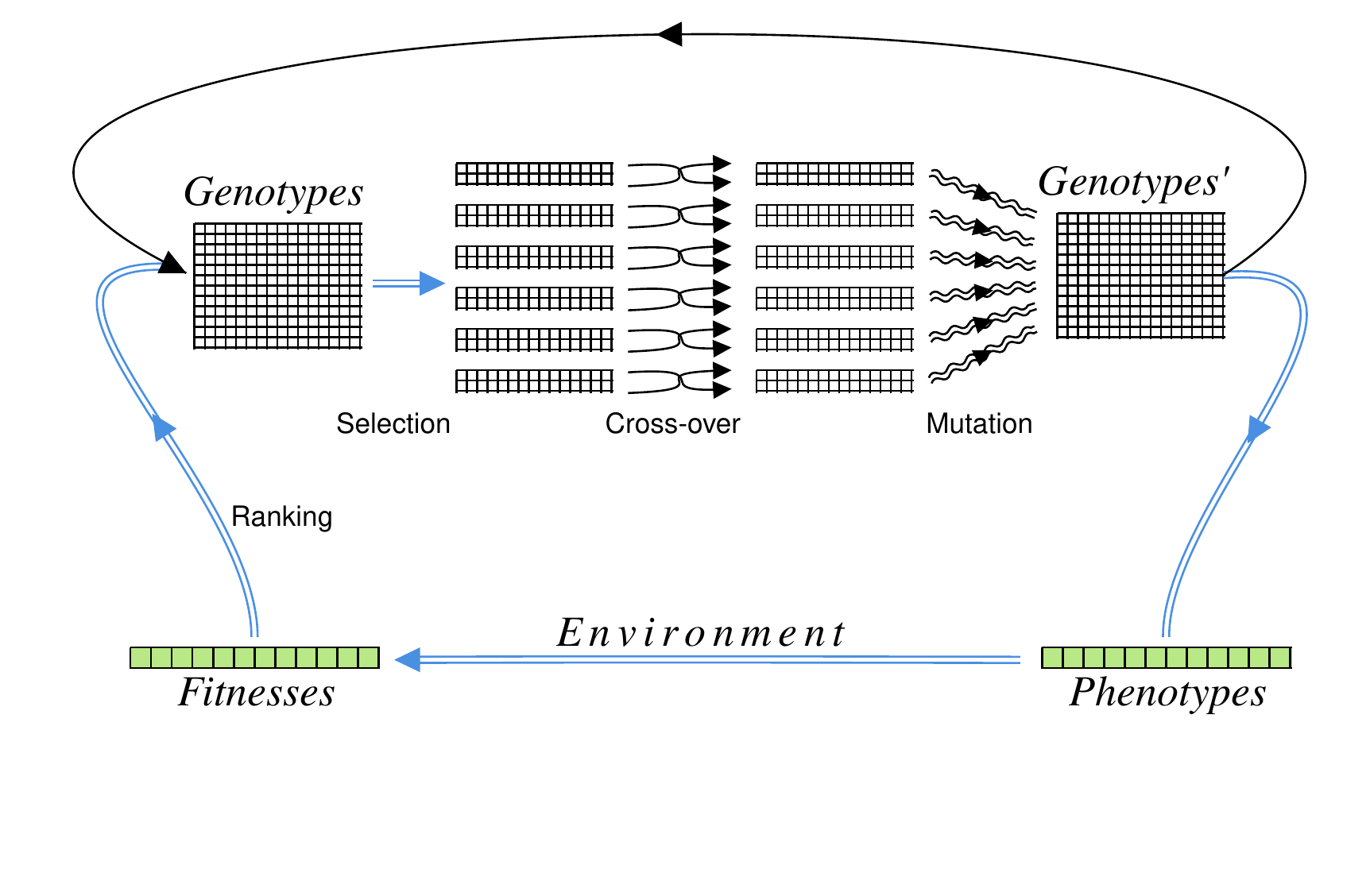}
 \vspace{-1.7cm}
\caption{The classical GA is a directed graph with two loops. The initial population of genotypes (on the right) give phenotypes which are in turn used to calculate the  fitnesses of the $P$ individuals in the `environment' which are then collected as a ranked population of parent genotypes (on the left). The fitness ranking of an individual determines the probability of it being selected to take part in a breeding pair. $P/2$ breeding pairs are formed in this manner, and then cross-over and mutation yield an entirely new generation, and the process is repeated. }
\label{fig:populationGA}
 \end{center}
\end{figure*}

Thus the procedure begins by forming a random population of $P$ individuals, by generating $P$ binary string genotypes with the appropriate number of entries, the so-called {\it alleles}, as shown in \fref{fig:individualGA}. To judge how successful a particular individual is we define a {\it fitness function}, which is some function of the emergent phenotype properties that increases monotonically as we get closer to solving the problem of interest. 

The population is then to be evolved using the three main ingredients specific to GAs: {\it selection, breeding} and {\it mutation}. The overall procedure is represented in \fref{fig:populationGA}, where we assume throughout that selection is based on the fitness-ranking.

{\it Selection} can proceed after having ranked the individuals according to their fitness. In \fref{fig:populationGA} we refer to this part of the algorithm, in which the fitnesses of the phenotypes are confronted with the problem of interest, as the {\it environment}. Individuals are then selected for breeding with a probability that increases with the ranking. A convenient choice is for the dependence to be linear, such that the probability $p_k$ of the $k^{\rm th}$ individual being selected for breeding is 
\begin{equation}
\label{eq:pk}
p_k~=~ \frac{2}{(1+\alpha)P} \left( 1+\frac{P-k}{P-1}(\alpha-1)\right)~,
\end{equation}
where $\alpha >1$ is a constant meta-parameter that can be thought of as the `learning rate' of the GA. In particular, the probability $p_1$ for the fittest individual to be selected is a multiple $\alpha$ of the probability $p_P$ for the least fit.
Typically, $\alpha$ is chosen in the range $2\leq\alpha\leq 5$.

{\it Breeding} is then performed with $P/2$ breeding pairs that have been selected in this manner, to form a new generation. It is implemented by cutting and splicing the pair at a number of matching random points. Typically a single point cross-over performs well enough, in which a cut is made at a single point and the `tails' swapped. 

{\it Mutation} is the final step, in which a small randomly selected fraction of the alleles in the newly formed generation is flipped. A typical value for the mutation rate is a few percent, but the optimum rate tends to be problem specific. 

As indicated in \fref{fig:populationGA}, this process is then repeated for multiple generations, typically a few hundred. If successful, the algorithm evolves the population until individuals begin to appear that correspond to viable solutions. 

The procedure described above can be thought of as the ``vanilla'' version of a GA, and many variations have been suggested (for a review of some of the improvements that can be made see Ref.~\cite{Abel:2021rrj}). However all implementations of GAs have in common  these three elements. One crucial aspect to remember for the later GQAA discussion is that  mutation is not just an improvement to the convergence, but is absolutely integral to the entire process because without it the system stagnates, as can be appreciated by optimising the mutation rate as in Ref.~\cite{Abel:2014xta} and below. 

\begin{figure}[!htbp]
 \vspace{-0.75cm}
 \hspace{-0.5cm}
\includegraphics[width=0.5\textwidth]{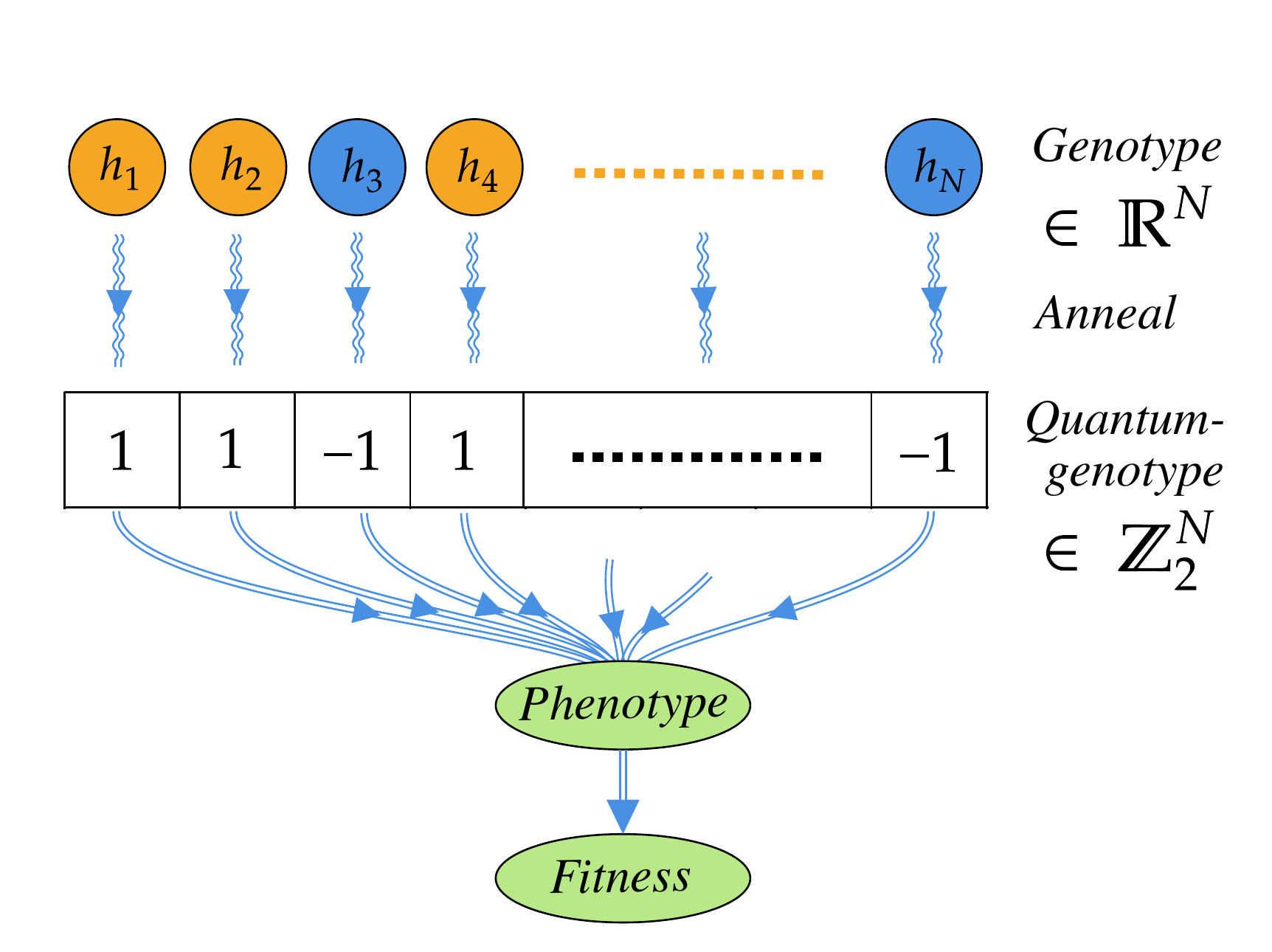}
 \vspace{-.4cm}
\caption{Representation of an individual member of the population in the GQAA, showing the relation between the genotype, quantum-genotype and phenotype. An individual corresponds to a chain of $N$ entries in the Ising model on the annealer. The genotype is defined classically in ${\mathbb R}^N$ and corresponds to the real biasing linear couplings for the individual, with orange nodes biasing positively (i.e. they have negative  $h$ values) and blue biasing negatively. A quantum-genotype lives in ${\mathbb Z}^N_2$ and is the corresponding discrete set of eigenvalues that is read from the annealer, which is influenced by the genotype, but also by couplings to the neighbouring members of the population. The quantum-genotype takes fluctuating values, with fitter individuals having larger modulus $|h|$ and hence enforcing their biasing more strongly, resulting in a form of `weighted mutation'. The phenotype is derived from the discrete quantum-genotype classically in the usual way.}
\label{fig:individual}
\end{figure}

\begin{figure}[!htbp]
 \vspace{-0.6cm}
 \hspace{-0.5cm}
\includegraphics[width=0.5\textwidth]{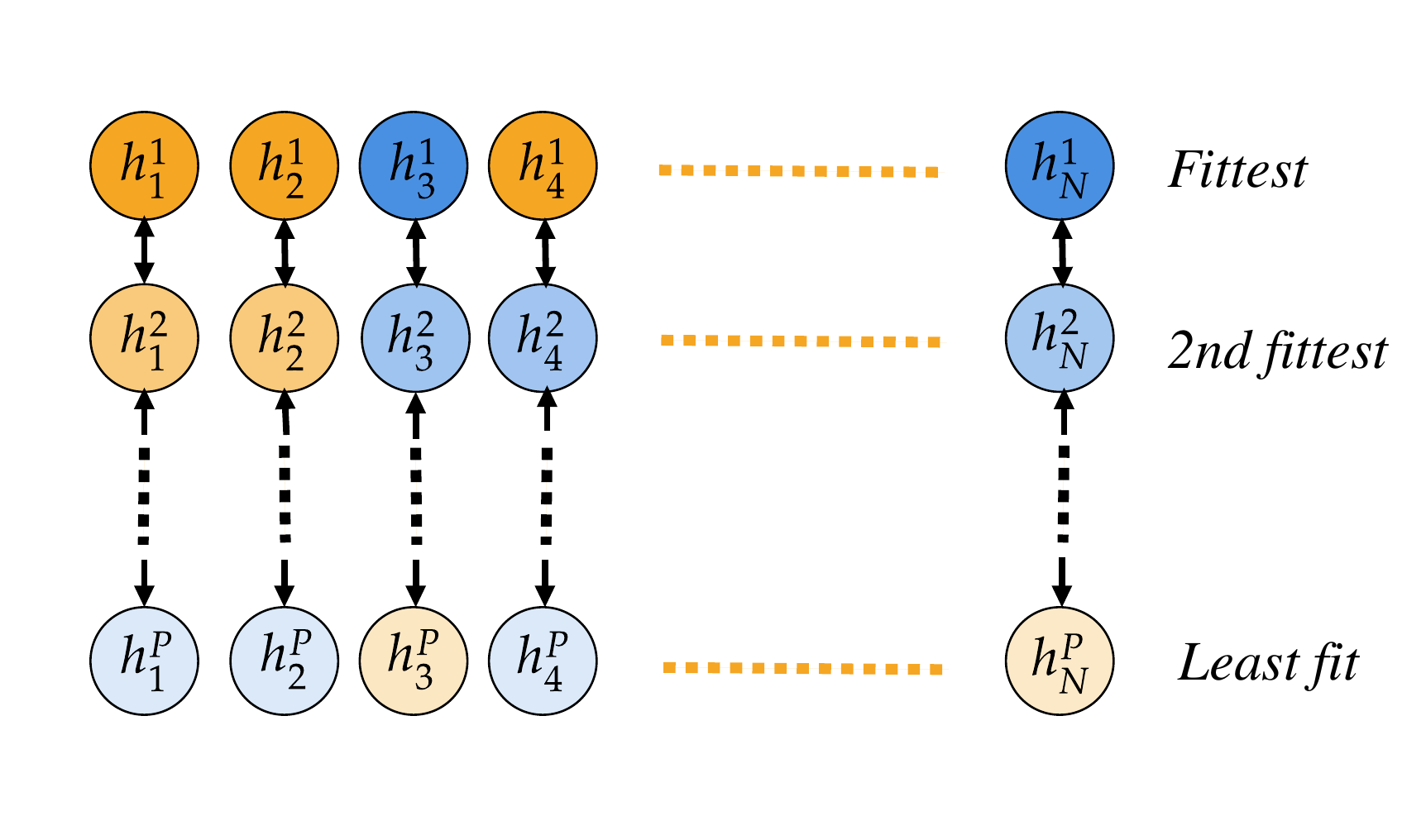}
 \vspace{-1.cm}
\caption{The arrangement of the population and the influence of fitness assignments before being selected for breeding. The genotype is stronger for the fitter individuals. Thus the quantum-genotype of the weaker individuals is influenced more by the couplings to the rest of the population (which are implemented vertically by allele). The Figure shows the na\"ive universal nearest-neighbour ferromagnetic configuration, in which the individuals with strong genotypes will more consistently enforce their corresponding quantum-genotypes, and will also impose them on neighbouring weaker members of the population, giving a form of `directed mutation'.}
\label{fig:population}
\end{figure}

\begin{figure*}[!htbp]
 \vspace{-.75cm}
\begin{center}\includegraphics[width=.85\textwidth]{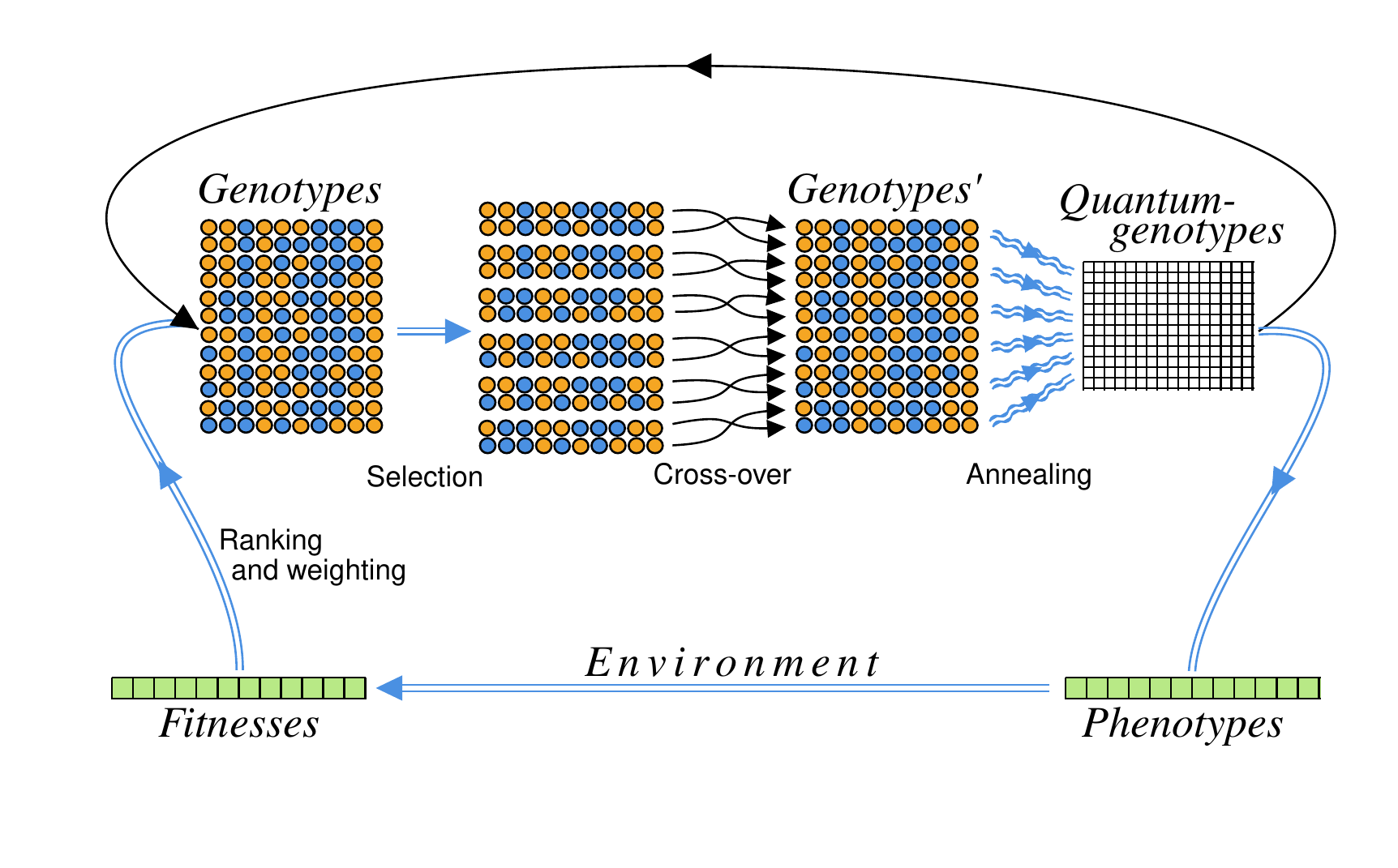}
 \vspace{-1.cm}
\caption{Like the classical GA in \fref{fig:populationGA}, the generic diagram for the QGAA has two loops. An initial population of discrete quantum-genotypes (on the right) is used to calculate the  fitnesses of the $P$ individuals in the `environment'. These are in turn used to form a fitness-ranked population of parent genotypes (on the left) whose continuous allele values are weighted by the fitness. As in the classical GA the fitness ranking of an individual determines the probability of it being selected to take part in a breeding pair. $P/2$ breeding pairs are formed in this manner, and then cross-over (keeping the weightings attached to the alleles) yields a new intermediate set of continuous genotypes. Completing the loops, quantum annealing yields a new set of quantum-genotypes and the process repeats.  }
\label{fig:populationGQAA}
\end{center}
\end{figure*}

\subsection{\label{Subsec:QA}Quantum Annealing}

Let us now describe the system that we wish to combine with the GA, which is a quantum annealer. Generally a quantum annealer works with a Hilbert space that is the tensor product of qubits, with a Hamiltonian of the form
\begin{align}
\label{eq:hami}
  \mathcal{H}(s) &~=~ A(s) \sum_\ell \sigma_{\ell,x}\\
    &\qquad+ \, B(s) \left(\sum_\ell h_\ell \sigma_{\ell,z} + \sum_{\ell m} J_{\ell m} \sigma_{\ell,z} \sigma_{m,z}\right) \, , \nonumber 
\end{align}
where $\sigma_{\ell,x}$ and $\sigma_{\ell,z}$ are the corresponding Pauli matrices acting on the $\ell^{\rm th}$ qubit, and $A(s)$, $B(s)$ are smooth functions such that $A(1) = B(0) = 0$ and $A(0) = B(1) = 1$, which are used to change the Hamiltonian during the anneal. The annealer can perform the following operations:
\begin{itemize}
  \item Set an initial state that is either the ground state of $\mathcal{H}(0)$ (known as \emph{forward-annealing}) or any eigenstate of $\bigotimes_\ell \sigma_{\ell, z}$ (\emph{reverse-annealing}).
  \item Fix the internal parameters $h_\ell$ and $J_{\ell m}$ of the Hamiltonian $\mathcal{H}(s)$.
  \item Allow the system to evolve quantum-mechanically while controlling $s$ as a piecewise-linear function $s(t)$ of time $t$, with $s(t_{\rm final}) = 1$, and $s(t_{\rm init}) = 0$ for forward-annealing, or $s(t_{\rm init}) = 1$ for reverse-annealing, where $t_{\rm init}$ and $t_{\rm final}$ are the initial and final times. The function $s(t)$ is called the \emph{annealing schedule}.
  \item Measure the observable $\bigotimes_\ell \sigma_{\ell,z}$ at $t = t_{\rm final}$.
\end{itemize}

Typically one chooses an anneal schedule such that the machine returns the ground state of the Ising-model Hamiltonian $\mathcal{H}(1)$.
Thus using the above operations one would usually attempt to encode the optimisation problem directly on the annealer, such that the global minimum of the  Hamiltonian corresponds to a solution. 
Thus in contrast with the GA the types of problems that can be treated in this traditional direct fashion are restricted to those that can be formulated as the minimisation of a quadratic Ising Hamiltonian which is a function $H$ of spin variables $\sigma_\ell = \pm 1$:
\begin{equation}
\label{eq:isingH}
 H(\sigma_\ell) ~=~ \sum_\ell h_\ell \sigma_\ell + \sum_{\ell m} J_{\ell m} \sigma_\ell \sigma_m~.
\end{equation}
To differentiate between the physical system and the embedded abstract problem, we will refer to the elements of the physical system ${\cal H}(\sigma_{\ell,z})$ as \emph{qubits}, and to the ones of the abstract system $H(\sigma_\ell)$ as \emph{spins}, where the classical spin values, $\sigma_\ell$, do not carry a $z$ index. 

We should make two clarifying remarks. First, the class of problems that can be treated can be extended by using `reduction' as described in Refs.~\cite{dattani:2019a,Abel:2022lqr,Abel:2021fpn,Abel:2022wnt}: namely given a high order polynomial in spins one can find a quadratic polynomial that it can be proved must share the same global minima. Second, in practice the annealer is not fully connected in couplings. At the time of writing the state of the art for such machines are those provided by D-Wave's~\cite{LantingAQC2017} \texttt{Advantage\_system4.1}, which has a so-called {\it Pegasus} structure that contains 5627 qubits but only 40279 couplings between them. One can overcome this problem using {\it chains}, in which qubits are locked together with strong couplings. Such chainlocking is usually done using an embedding programme, however this can use up many qubits on the machine if the Ising Hamiltonian is well connected, even when the problem is relatively small. 

\subsection{Genetic Quantum Annealing}

How might one incorporate some of the advantages of quantum annealers into a genetic algorithm? As is clear from the discussion in the previous subsection and in the introduction, from a quantum computing perspective the difficult part of the GA to encode would be the problem itself and the fitness function, namely the environment. Therefore our hybrid algorithm rests on continuing to treat the environment classically. 

The foundation of the approach is to redefine precisely what constitutes an `individual': our definition is shown in \fref{fig:individual}. In effect it adds another layer to the classical individual of \fref{fig:individualGA}.  Unlike the classical GA, the alleles in the genotype of an individual are {\it continuous}, and are comprised of a set of $N$ of the  linear couplings $h_\ell$ appearing in the spin Hamiltonian in Eq.~\eqref{eq:hami}, rather than a set of $N$ discrete alleles. The crucial feature that the quantum annealer introduces is the ability to convert this continuous genotype into a discrete genotype in a probablistic way by performing a quantum read of the corresponding spin eigenvalues. The latter discrete genotype we shall refer to as a {\it quantum-genotype}, due to its ability to take discrete but fluctuating values. Note that it is the quantum-genotype which will determine the phenotype and hence the fitness of the individual. This is the main ingredient provided by the quantum annealer, and we argue that this separation of the genotype from its physical manifestation quantises the crucial biological feature that has to be incorporated into classical GAs with an artificial mutation stage, namely the fact that the genetic code for an individual does not absolutely determine its phenotype.

For later reference it is also of course possible to define the discrete genotype that the $h_\ell$ couplings would {\it like } to enforce in the spins, which we refer to as the {\it classical-genotype}, which is simply 
\begin{equation}
\label{eq:sigmaclass}
    \sigma^{\rm cl}_\ell ~=~ -\, {\rm sign}(h_\ell)~. 
\end{equation}
Importantly, in the limit of zero quadratic couplings $J_{\ell m}$ on the annealer, and perfectly adiabatic annealing, the quantum-genotype is equal to the classical-genotype.

With this definition of an individual to hand, there is then a great deal of freedom in how one can configure them into a population on the annealer, and it is possible to go far beyond the classic framework in \fref{fig:populationGA}. The configuration that we adopt for the population is shown in \fref{fig:population}. There are two features that greatly enhance the classical GA. The first is that, being continuous, the genotype values do not need to have a universal weighting across the population, but they can be weighted by the fitness of the parent that gave rise to them. This weighting, which is shown in \fref{fig:population} as a stronger shading of the genotype nodes in the more highly ranked individuals, we will refer to as {\it nepotism}. For this study we will adapt a linear weighting for the couplings as follows:
\begin{equation}
\label{eq:nepotism}
|h_\ell| ~=~ \alpha_p \left(\frac{\alpha-1}{P-1} \, \ell + 1 \right) \, , \quad \ell = 0,...,P-1 \, .
\end{equation}

The second feature is that corresponding alleles across individuals in the population can be coupled in the Ising model (with (anti)ferromagnetic (positive)negative $J_{\ell m}$ couplings in Eq.~\eqref{eq:hami}). In other words, in contrast to a classical GA, the quadratic couplings in the quantum annealer allow the individuals to `see' the rest of the population. This allows highly ranked individuals to influence the quantum-genotypes of other members of the population, a feature we refer to as {\it quantum-polyandry}. To avoid clutter  \fref{fig:population} shows nearest-neighbour attractive (ferromagnetic) polyandry, but an important possibility to be discussed later is more connected quantum-polyandry, in which for example more members of a population can see the fittest individuals. 

The framework for the entire algorithm is shown in \fref{fig:populationGQAA}. Like the classical GA it is a two-loop directed graph, with the genetic material circulating in the top loop, while the feedback from the environment circulates in the bottom loop. The procedure begins as in the classical GA by randomly initialising the {\it discrete} quantum-genotypes in the population. The phenotypes and hence fitnesses for all the individuals are calculated, and the quantum-genotypes ranked accordingly. The biasing $h_\ell$ terms in Eq.~\eqref{eq:hami} on the quantum annealer are then filled with couplings of the opposite signs, and with moduli that grow according to these fitnesses, with the expectation that fitter individuals will be able to enforce their corresponding quantum-genotype more strongly. In addition polyandric $J_{\ell m}$ couplings are filled on the annealer (and kept fixed throughout) and then selection and breeding can be carried out in exactly the same manner as in the classical GA. However for the GQAA this entails swapping the continuous coupling parameters in the rows of the $h_\ell$ and  $J_{\ell m}$  matrices to form a new set of Ising model couplings. Finally this new Ising model is fed into the annealer to read off new sets of quantum-genotypes for the next generation. Note that if an individual is to breed several times then it may be preferable (and costs little) to avoid using the same set of  quantum-genotypes twice, but to collect a pool of multiple reads of the quantum-genotypes.

One notable aspect of this configuration is that in nepotism the fitness weighting follows clusters of `beneficial alleles' that may have conferred good fitness in the previous generation. From the viewpoint of Holland's original (and still somewhat controversial) schema theorem \cite{Holland1975} (see Ref.~\cite{Abel:2021rrj} for a summary of its various critiques), this can be thought of as a means of weighting powerful schema by the fitness of the individual from which they came.  

Meanwhile, as should be now be clear, it is the generation of the quantum-genotype by annealing which is playing the role that mutation played in the classical GA.
However in the configuration we are advocating here, both nepotism and quantum-polyandry work to direct the mutation towards configurations that confer fitness. 

\subsection{\label{subsec:polyandy} More on Quantum Polyandry}

The configuration of the polyandric couplings can take many forms, and the universal nearest-neighbour ferromagnetic couplings shown in Fig.~\ref{fig:population} are in fact not the optimal choice. Indeed this configuration leads to very rapid convergence and stagnation: in other words the fittest members of the population completely dominate the evolution very early. There are two modifications that can be made. One is to change the values of the couplings such that they can be either antiferromagnetic, or strengthened. In this work we will consider three values, namely ferromagnetic and antiferromagnetic couplings that are degenerate in modulus, and stronger ferromagnetic couplings. 

The second important modification is to allow more general topologies among the polyandric $J_{\ell m}$ couplings. Although this modification comes at the cost of additional qubits (as it requires higher order couplings than those available in the quadratic Ising model, which have to be implemented using chain-locking), the advantage is that it allows more individuals of a population to be influenced by the fittest ones. 

The motivation for these alternative configurations becomes clear when one considers also the possibility of antiferromagnetic couplings. One can imagine for example a stagnant situation where the fittest individuals are close to a solution modulo some minor flaw in the genotype. A set of antiferromagnetic couplings among the fittest subpopulation encourages some of its members to sacrifice their preferred quantum-genotype and explore minor modification. In this sense it is the GQAA equivalent of a {\it crowding penalty}. As such we expect the optimum proportion of antiferromagnetic couplings to be subdominant.

\begin{figure}[h]
\begin{center}
\includegraphics[keepaspectratio, width=0.5\textwidth]{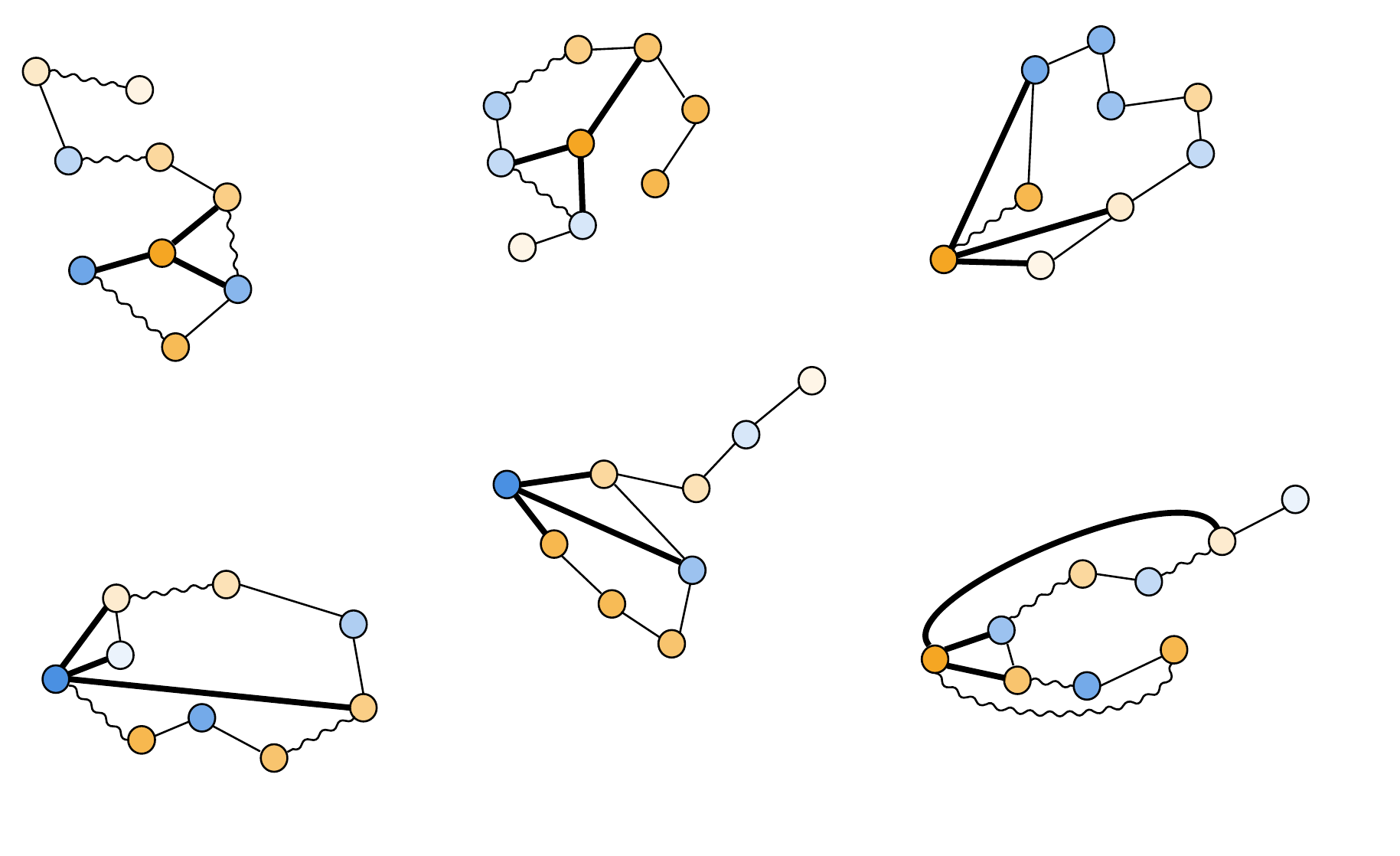}
\caption{The polyandric `islands' topology used for the analysis in this work, where each diagram indicates the connected islands repeated in every column of Fig.~\ref{fig:population}. Wavy lines refer to repulsive couplings, straight lines to attractive ones. Thicker lines indicate stronger attractive couplings between alleles belonging to fitter individuals and alleles belonging to weaker ones.}
\label{fig:spider_graph}
\end{center}
\end{figure}

Fig.~\ref{fig:spider_graph} is a example of such a non-trivial polyandric topology, which is in fact the configuration we use for this work. Each island of qubits corresponds to a single column in Fig.~\ref{fig:population}, with the genes of the fittest individual now attracting some of the genotypes of the weaker ones (thicker lines), instead of being coupled only to the second fittest creature as in Fig~\ref{fig:population}. Clearly there are a huge number of possible configurations, so we do not claim that this is the optimal one: it does however perform better than the trivial nearest-neighbour topology. 

\subsection{\label{subsec:polyandy} Practicalities: moving away from the classical GA limit}

Let now turn to an overview of the the practical implementation. First an important remark regarding the efficiency of the implementation is that the topology of the couplings in the Ising model remain constant throughout the QGAA, and only the values of the couplings are changed. This is crucial because, as already mentioned, finding a new embedding for the Ising model is done through an embedding algorithm which is itself computationally intensive. Thus all updates on the annealer are done by simply adjusting the classical $h$ and $J$ couplings. 

Next it is evident that the topologies in \fref{fig:populationGA}  and \fref{fig:populationGQAA} are the same, and indeed there is a limit in which the QGAA becomes isomorphic to the classical GA. This limit is when nepotism and polyandry are turned off, by setting all the polyandric $J_{\ell m}$ couplings to zero and by dialing up $s$ and making the fitness weighting universal, so that every individual just experiences the same level of mutation due to the quantum annealing stage. Thus the two parameters of polyandric coupling and fitness scaling are unique to the QGAA, while the parameter $s$ plays the role of the overall mutation rate. The fact that we can go parametrically to the classical GA allows direct comparison between the two approaches. 

Thus our approach will be to use the $h$ and $J$ couplings to move away from the classical GA limit, and this entails the use of a reverse anneal. The initialization of the reverse anneal is done with spins corresponding to the classical-genotype in Eq.~\eqref{eq:sigmaclass}. In the limit where the anneal parameter $s(t)$ is kept close to unity throughout the anneal, the spins do not change from their initialized values. By then turning on the quantum fluctuations but keeping the $J$ couplings set to zero, we can introduce a mild mutation rate in the spins, and the algorithm is effectively operating as a classical GA. Then dialing down $s$ and turning on $J$ for the various topologies allows one to move away from the classical GA. 

Some additional technical points: The annealer typically has an {\tt auto\_scale} parameter that automatically adjusts the $h$ and $J$ couplings to fill the physical range allowed on the annealer. This should be turned off to prevent the $h$ couplings being automatically scaled to large values, which would result in the spins being locked to their initialized state. 

In addition we implement {\it elitism} by replacing the least fit quantum-genotype with the best fit previous classical-genotype. We utilise the Pegasus annealer and find that a reverse anneal with $s(t)$ going to roughly $0.7$, and a schedule with ramp-up and ramp-down times of 10$\mu$s and total anneal times of 120$\mu$s, can reproduce a mutation rate in the fittest individuals that is similar to that of the classical GA (which as we shall see is optimised at a few percent for the problems we shall consider). However it should be noted that the mutation rate in the fitter individuals is less than that in the rest of the population precisely because their larger couplings impose their classical-genotypes more forcefully. It is also worth noting that the mutation rate is very sensitive to the minimum $s(t)$ value. 

To complete the description of the practicalities, we collate all of the variables and parameters that need to be considered, and the values we preferred for this study, in Table~\ref{table:params}.\\

\begin{table}[h]
\begin{center}
\resizebox{8.7cm}{!}
{
\begin{tabular}{| c | c | c |  }
 \hline
  \textbf{Parameter} & \textbf{Description} & \textbf{Value} \\ \hline \hline 
  {Topology} & Polyandric {$J_{\ell m}$ couplings} &  `Islands': {Fig.~\ref{fig:spider_graph}} \\
 \hline
 $\alpha $ & Learning rate, Eq.~\eqref{eq:pk}& 3 \\
 \hline
 $\alpha_p $ & Nepotism, Eq.~\eqref{eq:nepotism}& 0.05 \\
  \hline
 $\rho $ & Proportion of antiferromagnetic&  0.5 \\
  \hline
 $\rho' $ & Proportion of enhanced couplings & 0.064 ~|~ 0.06 \\
  \hline
 $\kappa $ & Strength of enhanced couplings & $-~\alpha\, \times \alpha_p$ \\
  \hline
 $s_q $ & Minimum anneal parameter & 0.74,0.72$ ~|~ $0.72,0.75,0.75  \\
  \hline
 $J_{ij} $ & Coupling strength & $\pm 0.07 ~|~ \pm 0.08$ \\
  \hline
\end{tabular}
}
\end{center}
\caption{Parameters, and the values used for this study. The two sets of values for $\rho$ and $\rho'$ correspond to the Diophantine problem on the left and function optimization on the right. The values of $s_q$ correspond to the consecutive $\kappa$ values on the left, and the three Taxicab numbers on the right.}
\label{table:params}
\end{table}\vspace{-0.7cm}

\section{Results for optimising a 2D function}

We now turn to the implementation of specific problems, and results. We should stress from the start that the goal of the exercise with the simple problems we shall look at is not to show that the GQAA can find results more quickly in real time: indeed the problems we will consider are so simple that a classical GA can work exceedingly fast. By contrast the goal is to show that the QGAA can find solutions by making far fewer `calls to the problem'. Minimising the number of times one has to evaluate the fitness of an individual is often the crucial efficiency factor in a GA when the problem to be solved is computationally intensive. Indeed, as mentioned in the introduction, the advantage of a GA is that it can operate on very complicated systems and this often means the majority of the effort is spent performing the computations required to find the fitnesses. It is precisely this advantage that the QGAA aims to enhance. In this sense the problems we will consider here should be considered as test problems. Thus the crucial parameter throughout will be the {\it call-count}, namely the number of individuals that have to be evaluated in total before a solution is found. We are also (in the spirit of optimisation of the  mutation rate of the classical GA in Ref.~\cite{Abel:2014xta}) looking for evidence that there is a preferred  configuration of the nepotism and the polyandric couplings. 

The first problem we shall consider is finding the global maximum of a complicated function. We will compare the GA and GQAA when perfoming the task of maximising the  function
\begin{align}
U_\kappa (x,y) ~\equiv~ & \frac{1}{2} \, \left(x(1-x) \,+ \, y(1-y)\right) \nonumber \\
& \qquad + \, 12 \cos(\kappa  xy)\sin(2x + y) \, ,
\label{eq:Uk}
\end{align}
in the region $(-4,4) \times (-4,4)$ with various values of $\kappa$ (increasing $\kappa$ introduces more local maxima making the function harder). For the fitness function we will just use the function itself. 

\begin{figure}
 \vspace{-1.cm}
\includegraphics[keepaspectratio, width=0.5\textwidth]{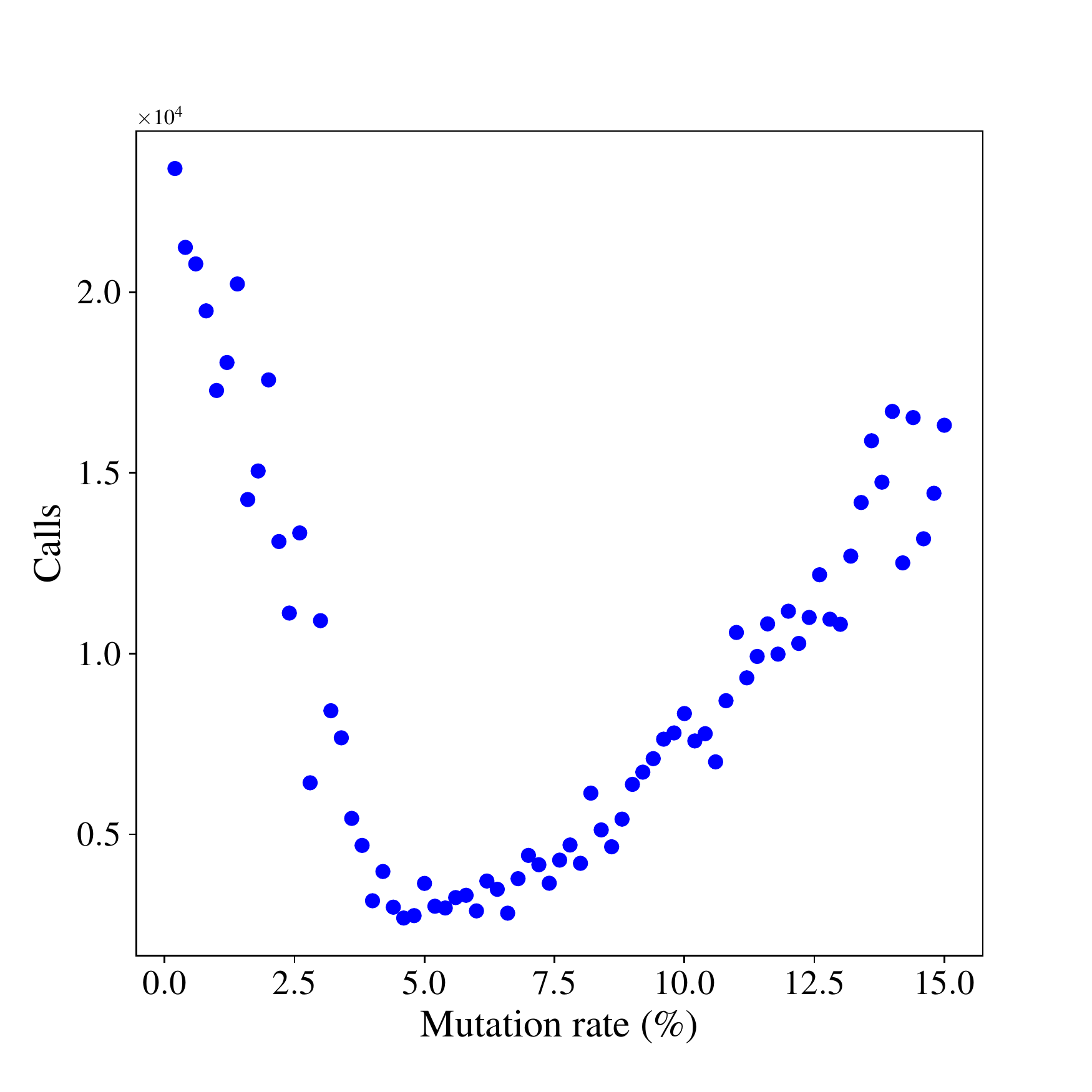}
\caption{Number of calls required to find a solution for different mutation rates for the classical GA. For this specific problem the best mutation rate is around $5 \%$.}
\label{fig:best_mut_rate}
\end{figure}

\begin{figure*}
 \vspace{-1.5cm}
\centering
\subfloat[][$\kappa = 1$]{
\includegraphics[width=0.5\textwidth]{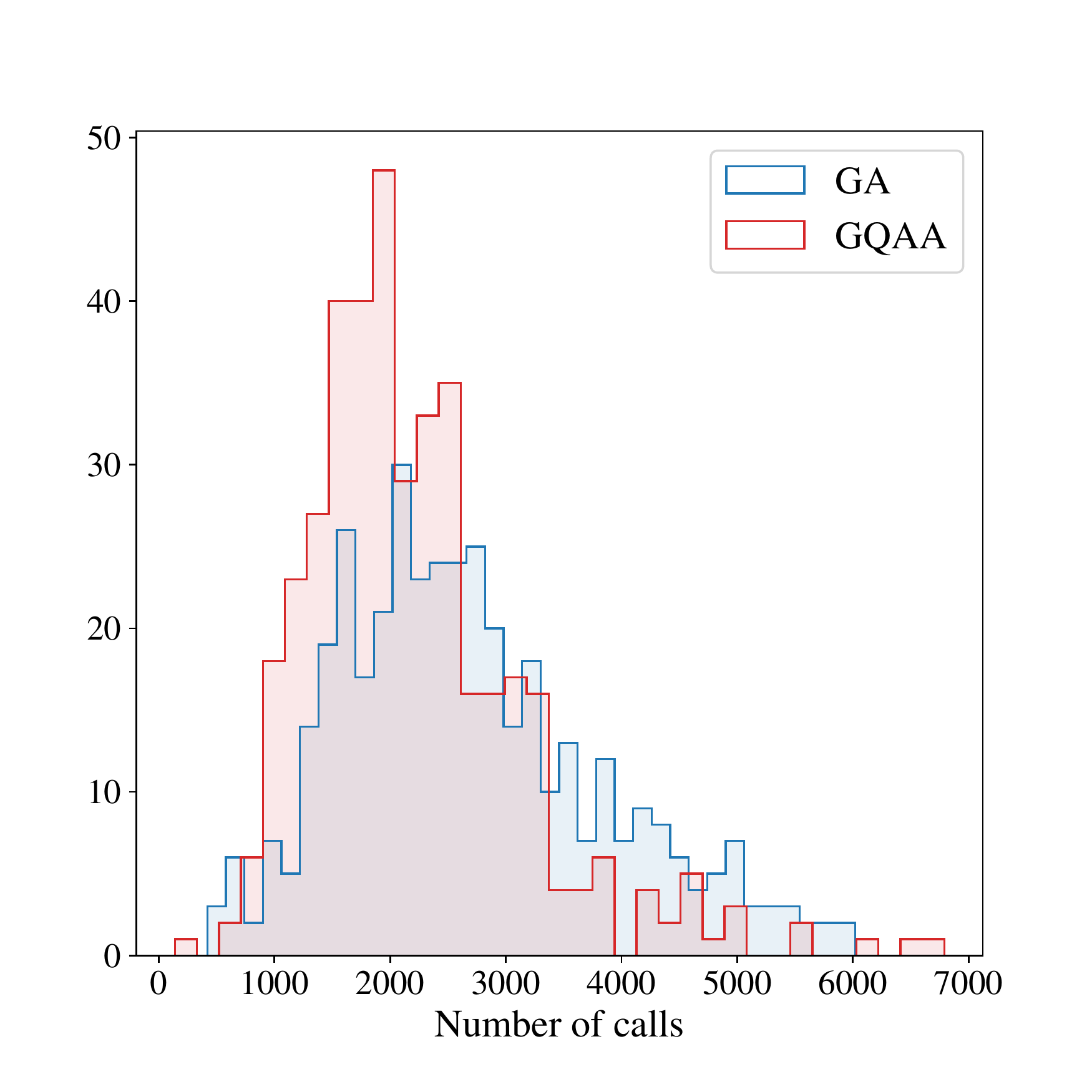}
\label{fig:hist_1}}
\subfloat[][$\kappa = 20$]{
\includegraphics[width=0.5\textwidth]{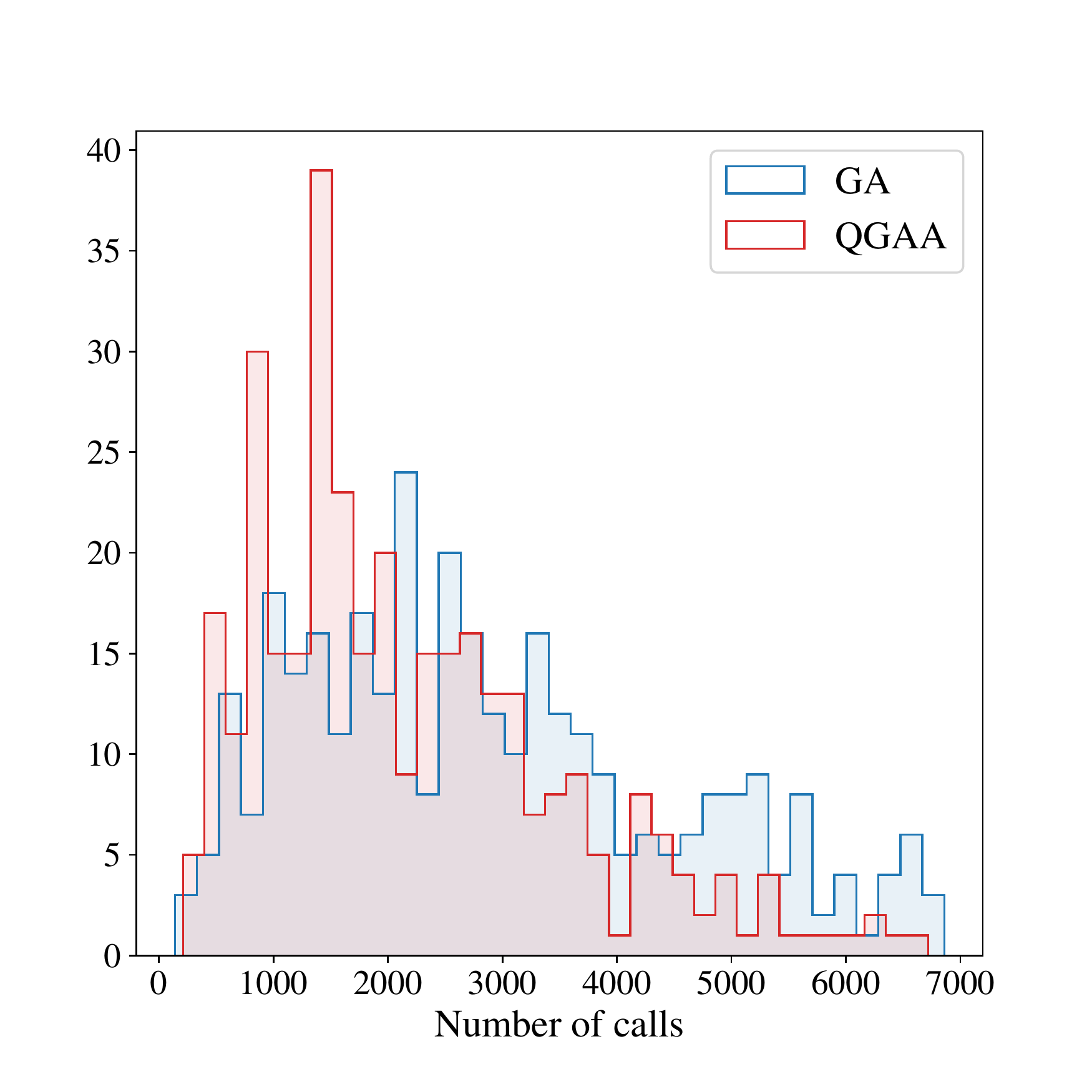}
\label{fig:hist_2}}
 \vspace{-0.1cm}
\caption{Number of calls required to find a solution with the demanded precision for finding the global maximum of the function in Eq.~\eqref{eq:Uk}. For $\kappa=1$ in Fig.~\ref{fig:hist_1} the average number of calls is: 2240 for GQAA and 2690 for GA (representing $\sim 18\%$ of improvement). Root-mean-square deviation (RMSD): 940.8 for QGAA and 1140.2 for GA. For the more complicated case with $\kappa = 20$ in Fig~\ref{fig:hist_2} the average number of calls is: 2186 for GQAA and 2883 for GA (representing $\sim 24\% $ of improvement). RMSD: 1331.6 for QGAA and 1620.2 for GA. In this second case, GA does not find any solution within the first $7000$ calls in $17.3\%$ of the cases. For GQAA this percentage reduces to $7.8\%$. In both cases the data are collected running 350 times for both GA and GQAA. The GQAA parameters are listed in Table~\ref{table:params}. }
\label{fig:hist}
\end{figure*}

We start with the easier case with $\kappa = 1$. According to Mathematica the function has a global maximum at $(x_{\text{max}},y_{\text{max}}) \approx (0.68708,0.170864)$ with $U_1 (x_{\text{max}},y_{\text{max}}) \approx 6.13506$.
In order to compare the GA and the QGAA we use a fractional binary representation with $2$ bits for the integer part and $10$ for the fractional one and $P = 70$, and define $(x_{\text{sol}},y_{\text{sol}})$ to be a solution if 
\begin{equation}
    U_1(x_\text{sol},y_{\text{sol}}) ~>~ 6.13503 \, .
\end{equation}
In order to make the best use of the GA, we optimise the mutation rate following Ref.~\cite{Abel:2014xta}, as shown in  Fig.~\ref{fig:best_mut_rate}. In Fig.~\ref{fig:hist_1} we plot the number of calls required to find the maximum of $U_1$ with the desired precision for both the GA and the GQAA.
Fig.~\ref{fig:hist_2} shows the same information for the function with $\kappa = 20$. In that case the maximum is in $(x_{\text{max}},y_{\text{max}}) \approx (0.488397,0.642488)$ with $U_{20} (x_{\text{max}},y_{\text{max}}) \approx 6.23257$. The lower bound that defines the solution is $6.23$.

It is clear that the advantage of the GQAA is significant. Moreover it is crucial to bear in mind that the simplicity of the problem somewhat constrains the possibility for the GQAA to give dramatic improvement. This is because a problem for which the search space is somewhat limited, and which can for example be treated by gradient descent, is already unlikely to yield great advantage for the GA over conventional search techniques. 

\vspace{-0.3cm}

\section{Results for Diophantine problems}

Thus the second problem we shall consider is somewhat harder, namely the solution of simple Diophantine problems, of the kind that were discussed in Ref.~\cite{Abel:2022wnt} also in the context of quantum annealers (but using direct Ising encoding of the problems). Such problems can be made arbitrarily hard. The specific problem on which we will test the GQAA method is the task of finding so-called ``Taxicab'' numbers, namely those numbers that can be expressed in more than one way as sums of equal powers. The most famous example is the number of Hardy and Ramanujan's eponymous taxi, ${\rm Ta}(2)=1729$, which is the smallest of the following list of numbers, all of which are expressible as the sum of two cubes in two different ways: 
\begin{align}
&1729 ~=~ 1^3 + 12^3 ~=~ 9^3 + 10^3, \nonumber\\
&4104 ~=~ 9^3 + 15^3 ~=~ 16^3 + 2^3, \nonumber \\
&20683 ~=~  24^3 + 19^3 ~=~ 10^3 + 27^3,  \nonumber\\
&...
\label{eq:taxi_solutions}
\end{align}
Following Ref.~\cite{Abel:2022wnt} we use the notation $(k,m,n)$, to refer to such numbers, where $k$ is the power, while $m$ and $n$ are the number of terms on each side, so that Ta$(2) ~=~ 1729$ is defined to be the smallest $(3,2,2)$ number, while Fermat's theorem is the statement that $(k,1,2)$ numbers only exist for $k=2$. Here we will test our methods on the computationally intensive $(3,6,6)$, $(3,7,7)$, $(3,8,8)$ numbers (examples of the former can be found in Refs.~\cite{Guy:1994a,Meyrignac,Eulernet,Weisstein,piezas}, 
while to our knowledge the latter first appeared in 
Ref.~\cite{Abel:2022wnt}). 

In order to treat the problem on a GA or QGAA for the
$(3,n,m)$ numbers, where $n,m \in \mathbb{N^+}$ we use the following fitness function:
\begin{equation}
\tilde{f} ~=~ - \left(\sum_{i=1}^n a_i^3 - \sum_{i=1}^m b_i^3\right)^2 + \mbox{\it constraints} \, , 
\label{eq:fitness2}
\end{equation}
where the constraints refer to an additional Kronecker-delta penalty if any of the $a_i$ are equal to any of the $b_i$.

Optimising the mutation rate for the GA, as for the previous problem, we begin by  
displaying some specific examples of solution-finding for the (3,6,6) and (3,8,8)  problems. Parameterising the integers with the obvious 5 digit binary encoding, we find the examples shown in Table~\ref{table:solutions}. It is clear from these examples that the GQAA  consistently  outperforms the classical GA in finding solutions. 

\begin{table*}[!]
\vspace{-0.2cm}
  \centering
 \resizebox{17.8cm}{!}
{
\subfloat[][GA: (3,6,6)]
{
\begin{tabular}{| c | c | c | c|   }
 \hline
  \textbf{Call-count} & \textbf{Gen's} & \textbf{~P~} &\textbf{Solution}\\
  \hline \hline 
91650 & 1833 & 50 & $ (~1,  3,  3,  9, 13, 11 ~|~ 12,  8, 4,  6,  4, 12~) $ \\
61320 & 876 &  70 &  $ (~12,29, 31, 4, 5, 6~|~ 9, 28, 23, 18, 25, 2~) $ \\
9270 & 309 &  30 &  $ (~27, 15, 27, 18, 10, 27 ~|~ 30, 16, 12, 31, 12, 17~) $ \\
21390 & 713 &  30 &  $ (~2, 2, 21, 27, 15, 15~|~ 30, 8, 3, 14, 11, 16 ~) $ \\
25680 & 856 &  30 &  $ (~7, 3, 31, 16, 24, 20 ~|~ 28, 18, 1, 8, 28, 18~) $ \\
11610 & 387 &  30 &  $ (~15, 26, 15, 15, 15, 22~|~ 19, 10, 2, 21, 27, 17~) $ \\
\hline
\end{tabular}
\label{table:GAsolutions}}
\subfloat[][GQAA: (3,6,6)]
{
\begin{tabular}{| c | c | c | c|   }
 \hline
  \textbf{Call-count} & \textbf{Gen's} & \textbf{~P~} &\textbf{Solution}\\
  \hline \hline 
2130 & 71 & 40 & $ (~13,9,8,6,8,8 ~|  ~10,2,10,11,2,11~) $ \\
2680 & 67 &  40 &  $ (~25,9,5,10,3,9 ~|  ~7,15,21,12,11,13~) $ \\
8520 & 284 &  30 &  $ (~18,15,29,27,23,13 ~|  ~16,26,11,30,26,4~) $ \\
2220 & 74 &  30 &  $ (~1,23,29,18,9,13 ~|  ~24,28,10,3,20,8~) $ \\
3060 & 102 &  30 &  $ (~21,29,19,3,11,3 ~|  ~26,13,14,9,22,20~) $ \\
1170 & 39 &  30 &  $ (~24,26,25,3,7,11 ~|  ~22,20,20,28,5,1~) $ \\
\hline
\end{tabular}
\label{table:GQAAsolutions}}
\label{table:solutions}}

 
\resizebox{17.8cm}{!}{
\subfloat[][GA: (3,8,8)]
{
\begin{tabular}{| c | c | c | c|   }
 \hline
  \textbf{Call-count} & \textbf{Gen's} & \textbf{~P~} &\textbf{Solution}\\
  \hline \hline 
25160 & 629 & 40 & $ (~22,  6,  2,  2, 19, 24, 4, 24 ~|~ 21, 13,  13, 5,  5,  12, 3, 31~) $ \\
8280 & 207 &  40 &  $ (~26,  6,  3, 13, 10, 19, 18, 26 ~|~ 17,  7,  5, 29, 16, 25,  4, 12~) $ \\
21440 & 536 &  40 &  $ (~2, 20, 23, 22, 15,  2, 22, 1  ~|~ 4,  3, 10, 29, 11,  5,  7, 26~) $ \\
24640 & 616 &  40 &  $ (~16, 16,  8, 21, 24, 31, 9, 21 ~|~ 30, 14, 19, 19, 29, 15,  7,  1~) $ \\
30280 & 757 &  40 &  $ (~21, 10, 15, 14, 19, 24, 22, 26 ~|~ 17,  2, 20,  7,  4,  6, 31, 28~) $ \\
26800 & 670 &  40 &  $ (~13, 21, 14, 17, 19, 27, 14, 15 ~|~ 12,  9, 26,  8,  2, 22, 28,  7~) $ \\
\hline
\end{tabular}
\label{table:388_GAsolutions}}
\subfloat[][GQAA: (3,8,8)]
{
\begin{tabular}{| c | c | c | c|   }
 \hline
  \textbf{Call-count} & \textbf{Gen's} & \textbf{~P~} &\textbf{Solution}\\
  \hline \hline 
6800 & 170 & 40 & $ (~25, 13, 21,  3, 25, 1, 25, 21 ~|~ 11, 24, 20, 30, 14, 14, 22, 11 ~) $ \\
4440 & 111 &  40 &  $ (~17, 29, 21, 14, 1, 19, 17,  1 ~|  ~26, 30,  7,  3, 11, 24, 12,  8~) $ \\
3680 & 92 &  40 &  $ (~24, 28,  8, 25, 20, 12, 6, 10 ~|  ~16, 23,  4, 26, 19, 16, 23, 18~) $ \\
5280 & 132 &  40 &  $ (~20, 30,  3, 18, 14, 12, 12, 30 ~|  ~ 25,  5,  2, 21, 19, 26, 29~) $ \\
2720 & 68 &  40 &  $ (~7, 28,  6,  4,  6, 28, 15, 29 ~|~ 19, 26, 21, 10,  9, 18, 25, 25~) $ \\
3760 & 94 &  40 &  $ (~17,  9, 16, 23,  4, 21, 17, 14 ~|~ 10,  5, 26,  7, 27,  3,  5,  2 ~) $ \\
\hline
\end{tabular}
\label{table:388_GQAAsolutions}}
\label{table:solutions}}

\caption{
Hands-on with the GQAA versus the classical GA for finding (3,6,6) and (3,8,8) Taxicab numbers for integers up  to 32. Thus the search space is in principle $2^{60}\approx 10^{18}$ for the (3,6,6) problem, and $2^{80}\approx 10^{24}$ for the (3,8,8) problem. However the fact that the GA works with a rather small population indicates that the number of solutions in the search space is large.  Nevertheless the efficiency of the GQAA is still consistently up to an order of magnitude better in terms of the call-count.}
\label{table:solutions}
\end{table*}

\begin{figure*}[h]
\vspace{-2cm}
\centering
\subfloat[][Taxicab (3,6,6)]{
\includegraphics[width=0.45\textwidth]{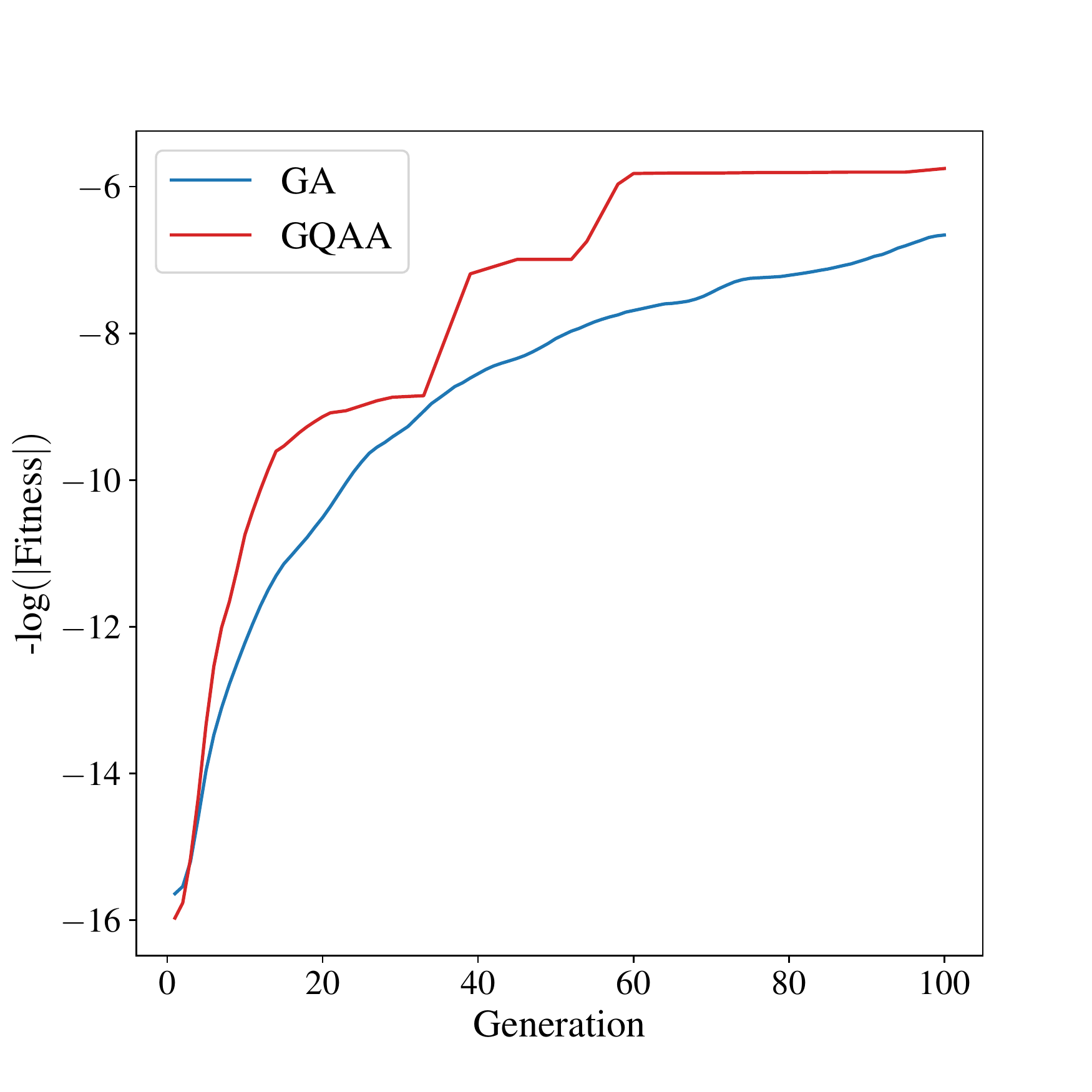}
\label{fig:fitness_vs_generations_1}}
\subfloat[][Taxicab (3,7,7)]{
\includegraphics[width=0.45\textwidth]{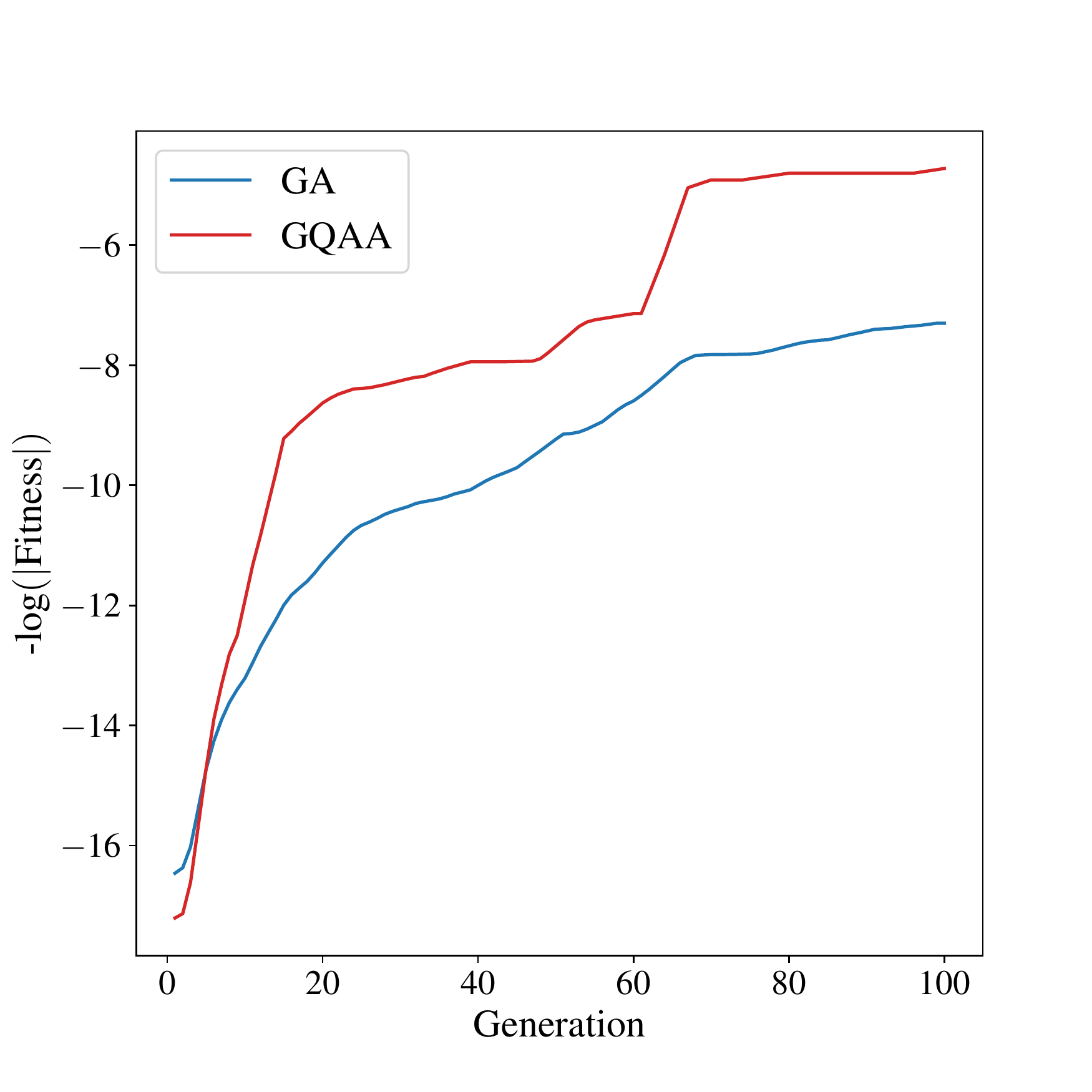}
\label{fig:fitness_vs_generations_2}}

\subfloat[][Taxicab (3,8,8)]{
\includegraphics[width=0.45\textwidth]{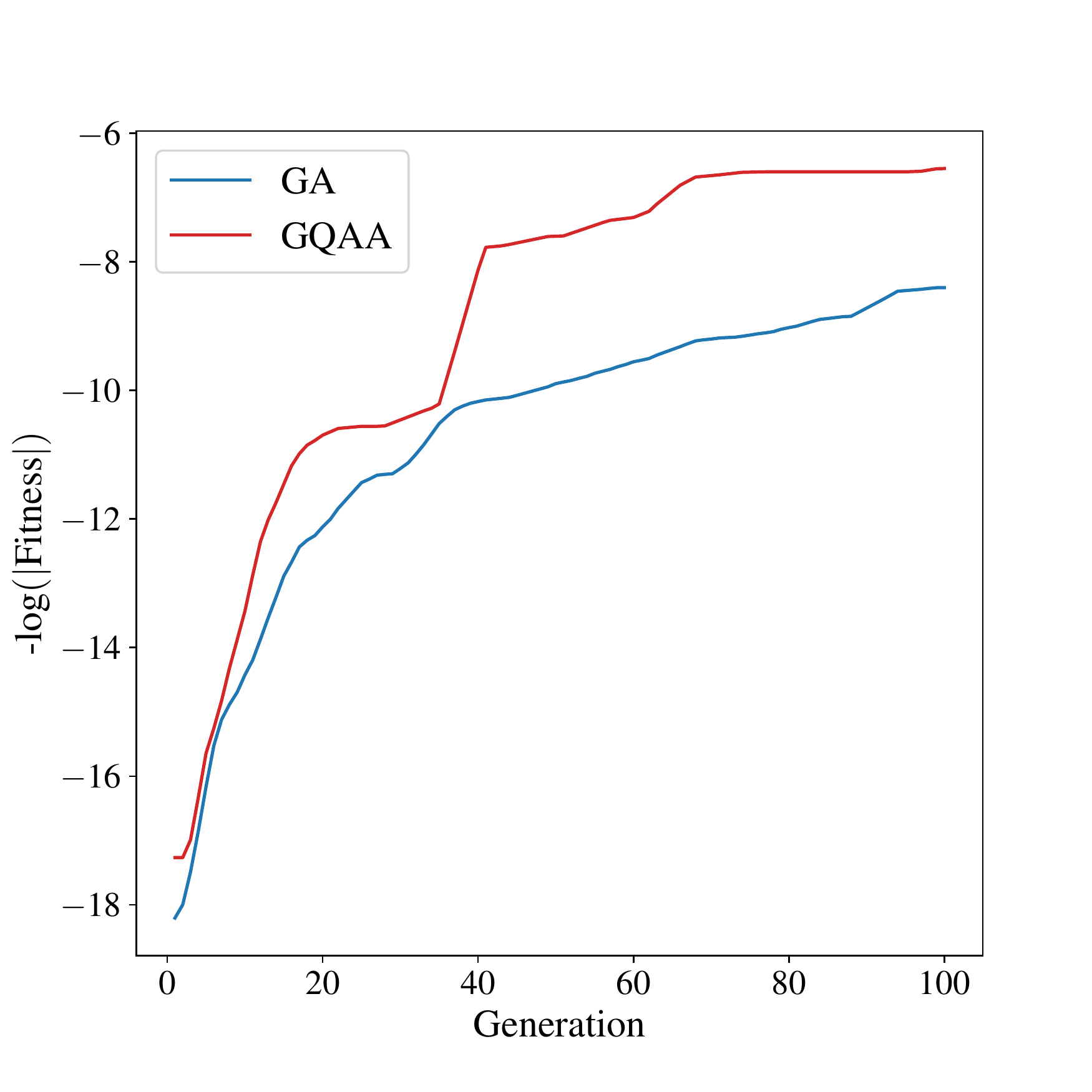}
\label{fig:fitness_vs_generations_3}}
\caption{Averaged fitness of the fittest creature for both GA and GQAA throughout the first 100 generations for $(3,n,n)$ Taxicab problems. For the (3,6,6) problem, after 100 generation the fittest creature has, on average, fitness -759.1 in the GA case and -312.4 in the GQAA one. For the (3,7,7) case in Fig.~\ref{fig:fitness_vs_generations_2} the fittest creature has, on average, fitness -1445.2 in the GA case and -108.3 in the GQAA one, while in for the (3,8,8) case in Fig.~\ref{fig:fitness_vs_generations_3} the fittest creature has, on average, fitness -4383.3 in the GA case and -655.44 in the GQAA one.}
\label{fig:fitness_vs_generations}
\end{figure*}

As the solutions to these problems are harder to find, we performed a systematic comparison by studying the progress of the maximum fitness in the population throughout the evolution. Figure~\ref{fig:fitness_vs_generations} 
compares the progress for the $(3,n,n)$ problems, where the maximum fitness has been averaged over 50 trials with $\beta = 5$ and $P = 30$, with the other GQAA parameters as given in Table.~\ref{table:params}.

Again the performance appears to be roughly an order of magnitude better in terms of the maximum fitness for the GQAA versus the GA. It is notable that the GQAA appears to behave somewhat differently in order to achieve increasing fitness. It evolves via a series of notably more dramatic jumps in improvement than the GA -- for a particular choice of parameters the jumps in fitness in the GQAA persist (it is not a statistical artefact), whereas the GA appears to progress more smoothly but more slowly overall. 

\section{\label{Sec:Conclusions}Conclusions}

We have presented a method, the GQAA, to implement  genetic algorithms on quantum annealers, which takes advantage of the quantum properties of the annealers to significantly improve performance over a classical GA. 

There are two novel aspects of the GQAA with respect to the classical GA. The first is that the genotype of each individual is encoded as a continuous set of annealer couplings and not as a discrete set. This then yields via the annealer the discrete `quantum-genotypes' from which the individuals are actually to be built, as shown in the scheme in \fref{fig:populationGQAA}. Thus there are several ways in which the genotype couplings can be chosen to influence their offspring. For example the fitness of previous generations can be encoded in the couplings so as to enforce good schema in the next generation (adopting Holland's viewpoint) -- so-called {\it nepotism}. In a classical GA such enhancement can only come at the expense of large portions of the population carrying the same schema. The second novel feature that is incorporated into the GQAA is so-called {\it polyandry}. That is the Ising couplings on the quantum annealer can be turned on (with various topologies and couplings) so that the entire population can be influenced by the fittest individuals. 

In the examples we discussed we found that the QGAA performed up to roughly an order of magnitude better, in terms of the number of individuals that had to be constructed before a solution of some simple test problems was found. It is worth noting that this improvement was found with rather generic choices of parameters, and a full optimisation of the method was not carried out in this study (due to time constraints on the annealer). It would be naturally very interesting to perform a full study to determine the optimum set-up for performance. \\ \vspace{-0.1cm}


\noindent {\it {Acknowledgements}:}  We would like to thank  Nick Chancellor, Andrei Constantin, Juan Criado, Thomas Harvey and Andre Lukas for helpful discussions. S.A. is supported by the STFC under grant ST/P001246/1. 

\bibliographystyle{inspire}
\bibliography{references,referencesSAMS}

\end{document}